\documentclass[aps,pra,reprint,twocolumn,floatfix]{revtex4-2}
\pdfoutput=1
\usepackage[utf8]{inputenc}
\usepackage[english]{babel}
\usepackage[T1]{fontenc}
\usepackage{amsmath}
\usepackage{hyperref}
\usepackage{adjustbox}
\usepackage{verbatim}
\usepackage{amsfonts, amsmath, amssymb}
\usepackage{graphicx}
\usepackage{dcolumn}
\usepackage{bm}
\usepackage[normalem]{ulem}
\usepackage{color}
\usepackage{braket}
\usepackage{physics}
\usepackage{dsfont}
\usepackage[caption=false]{subfig}
\usepackage{color}
\usepackage{soul}
\usepackage{mathtools}
\usepackage{float}
\usepackage{booktabs}
\newcommand{\be}{\begin{equation}}
\newcommand{\ee}{\end{equation}}
\newcommand{\ba}{\begin{eqnarray}}
\newcommand{\ea}{\end{eqnarray}}

\newcommand{\ssx}{\hat{\sigma}^x}

\newcommand{\ssz}{\hat{\sigma}^z}

\newcommand{\thetat}{{\tilde{\vb*{\theta}}}}
\newcommand{\param}{\theta} % single model parameter
\newcommand{\bias}{b} % single model bias
\newcommand{\params}{\boldsymbol{\param}} % model parameters
\newcommand{\biases}{\boldsymbol{\bias}} % model biases
\allowdisplaybreaks[1]

\def\XXint#1#2#3{{\setbox0=\hbox{$#1{#2#3}{\int}$}
    \vcenter{\hbox{$#2#3$}}\kern-.5\wd0}}

\DeclareMathOperator*{\argmin}{\arg\min}

\newcommand{\sket}[1]{\ket{\smash{#1}}}

\newcommand{\pidx}[1]{{\mbox{\tiny $(#1)$}}}

\newcommand{\kd}[1]{\textcolor{black}{#1}}
\usepackage{tikz}
\usepackage{lipsum}

\begin{document}

\title{Dynamics with autoregressive neural quantum states: application to critical quench dynamics}

\date{\today}

\author{Kaelan Donatella}
\author{Zakari Denis}
\author{Alexandre Le Boit\'e}
\author{Cristiano Ciuti}
\affiliation{Universit\'{e} Paris Cit\'{e}, CNRS, Mat\'{e}riaux et Ph\'{e}nom\`{e}nes Quantiques, F-75013 Paris, France}

\begin{abstract}
  Despite very promising results, capturing the dynamics of complex quantum systems with neural-network ansätze has been plagued by several problems, one of which being stochastic noise that makes the dynamics unstable and highly dependent on some regularization hyperparameters. We present an alternative general scheme that enables one to capture long-time dynamics of quantum systems in a stable fashion, provided the neural-network ansatz is normalized, which can be ensured by the autoregressive property of the chosen ansatz. We then apply the scheme to time-dependent quench dynamics by investigating the Kibble-Zurek mechanism in the two-dimensional quantum Ising model. We find an excellent agreement with exact dynamics for small systems and are able to recover scaling laws in agreement with other variational methods.
\end{abstract}

\maketitle

\section{Introduction}

The use of artificial neural networks to represent wavefunctions has opened up a new avenue in the understanding of many-body quantum systems~\cite{Carleo_2017}. These neural quantum states (NQS) have found many important applications, including finding the ground state~\cite{Sharir2020, Choo_2019}, investigating the dynamics of strongly correlated systems~\cite{Schmitt_2020,gutierrez2021real}, quantum tomography~\cite{Torlai_2018}, open quantum systems~\cite{Vicentini_2019, Hartmann_2019, Yoshioka_2019, Nagy_2019, Vicentini_2022} and the classical simulation of quantum circuits~\cite{Jonsson2018, Medvidovic_2021}. In fact, NQS approaches have proved to be the most accurate variational method in approximating the $J_1$--$J_2$ model's ground state in the frustrated regime~\cite{Choo_2019,nomura2021,roth2023,chen2023}. In addition, several works have recently demonstrated the superior capacity of some neural-network architectures over tensor-network states in representing volume-law entangled states~\cite{deng2017,glasser2018,Levine_2019, sharir2022} or area-law entangled states in 2D~\cite{Wu_2022}. \kd{This property makes NQS very promising for numerically simulating the dynamics of quantum many-body systems, since their growth of entanglement with time ~\cite{eisert2006,bravyi2007,marien2016} is a limitation for tensor-network approaches.}%All these results indicate that NQS methods are very promising to numerically describe quantum many-body systems.

The study of nonequilibrium dynamics is essential to understand spectral properties of complex quantum systems and investigate, for instance, correlation propagation~\cite{Cheneau_2012}. Impressive results for the transverse-field Ising model on large lattices have been obtained using convolutional neural networks~\cite{Schmitt_2020}. It is therefore of particular interest to continue investigating such problems that could yield important insights into nonequilibrium phenomena. In particular, studying the quantum Kibble-Zurek mechanism in higher dimensions \kd{has recently attracted attention}~\cite{Schmitt_2021} % has remained relatively untouched~\cite{Schmitt_2021}, 
and NQS techniques are deemed to be employed for the dynamics of time-dependent finite-size systems.

While early works focused on the restricted Boltzmann machine (RBM) ansatz, more recent works on ground-state search have employed networks that are closer to the state of the art in machine learning such as autoregressive convolutional models~\cite{Sharir2020} or recurrent neural networks~\cite{Allah2020, Hibat_Allah_2021}, for which the accuracy of the variational ground state energy was significantly improved. These networks have a so-called autoregressive structure, which means that one can perform direct sampling of uncorrelated configurations for arbitrarily large system sizes, thereby reducing the number of required samples. It is therefore of crucial importance to investigate the use of similar networks for quantum dynamics.

However, the widespread application of more complex and autoregressive neural networks for quantum dynamics has been held back by issues arising with time-dependent variational Monte Carlo (t-VMC)~\cite{becca_sorella_2017}. This procedure, most often used to propagate NQS over time, involves the inversion of a stochastically constructed singular matrix, which makes it particularly prone to noise. While some regularization techniques have helped improving the accuracy of the method~\cite{hofmann2021role,Schmitt_2020}, accessing all regimes at long times via t-VMC remains a challenge~\cite{Park2020,dawid_2022,netket_2021}.

In this work, we show that the stability of t-VMC strongly depends on the chosen ansatz, and that, in particular, it fails when applied to those based upon recurrent neural networks (RNN). To circumvent this issue, we propose an alternative scheme to numerically solve the dynamics of quantum systems. The scheme consists in casting an arbitrary Runge-Kutta integration scheme of any order into minimizing a variational distance at each time step, while only involving a polynomial overhead. This enables one to employ recurrent neural networks for quantum dynamics, which leads to a drastic reduction in the number of required samples. Our scheme can be implemented for any order of a chosen integration method, for a polynomial overhead in memory. We then apply our scheme to both time-dependent and sudden quenches, enabling us to recover Kibble-Zurek scaling laws for large system sizes and high precision on the dynamics during the full quench. 

\begin{figure*}
    \centering
    \includegraphics[width=\textwidth]{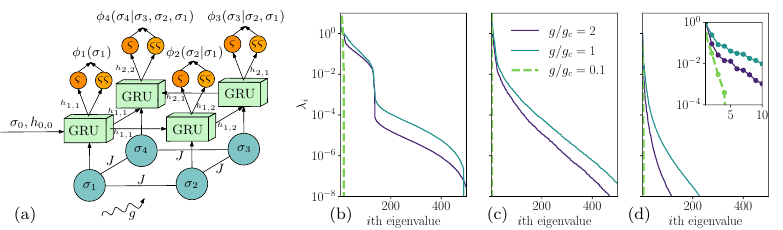}
    \caption{(a): Scheme of the autoregressive neural quantum state (NQS) ansatz used throughout the paper, for the specific simple example of $N = 4$ spins with open boundary conditions. The system is governed by the transverse field Ising (TFI) Hamiltonian, and configurations are fed into a gated recurrent unit (GRU) in an autoregressive fashion, each cell being fed with a hidden vector $\vb*{h}_{i,j}$ and a spin $\sigma_i$. Note that the variational parameters (matrices $\vb*{W}$ and vectors $\vb*{b}$), defined in Appendix \ref{app:a}, are the same for each GRU. The way the $\vb*{h}_{i,j}$ are generated is also reported therein. After the GRU cell, a softmax ($\varsigma$) and a softsign ($\varsigma \varsigma$) transformations are applied to obtain the amplitudes and the phases of the conditional amplitudes $\phi_i(\sigma_i|\sigma_{i-1},\ldots,\sigma_1)$. The ansatz wavefunction is just the product of all the $\phi_i$ functions. (b)--(d): Spectra of the quantum geometric tensor $\vb{S}$ for different ansätze optimized to the ground state of the TFI model, for various values of $g/J$. Panel (b) is for a complex restricted Boltzman machine (RBM) ansatz with $\alpha = 2$, panel (c) for a three-layer convolutional autoregressive network with complex parameters, and panel (d) for a GRU ansatz with real parameters and $d_h = 10$. For this last panel, the spectrum of the imaginary part of the $\vb{S}$ matrix is plotted, as this is the quantity to invert to simulate dynamics for ansätze with real parameters.
    }
    \label{fig:scheme_and_SR}
\end{figure*}

\section{Time-evolution of Neural Quantum States}
\subsection{Autoregressive neural quantum states}
Let us define an autoregressive NQS by its complex amplitudes $\psi_{\vb*{\theta}}(\vb*{\sigma}) = \braket{\vb*{\sigma}}{\psi_{\vb*{\theta}}}$, with $\ket{\vb*{\sigma}} = \ket{\sigma_1 \sigma_2 \ldots \sigma_N}$ the basis elements of the Hilbert space $\ket{\psi_{\vb*{\theta}}}$ belongs to, $\sigma_i$ local spin configurations and ${\vb*{\theta}}$ the variational parameters. The amplitudes $\psi_{\vb*{\theta}}(\vb*{\sigma})$ satisfy the \textit{autoregressive} property:
\begin{equation}
    \psi_{\vb*{\theta}}(\vb*{\sigma}) = \Pi_{i}^{N} \phi_i(\sigma_i|\sigma_{i-1} \ldots \sigma_1),
\end{equation}
with $\phi_i$ some normalized parametrized functions, which we refer to as the conditional amplitudes\footnote{Depending on the chosen ansatz, the conditional amplitudes will depend on only a subset of the variational parameters $\vb*{\theta}$.}. It follows that the Born conditional probabilities, $p_i(\sigma_i|\sigma_{i-1} \ldots \sigma_1) = |\phi_i(\sigma_i|\sigma_{i-1} \ldots \sigma_1)|^2$, share this structure. Furthermore, upon ensuring they are properly normalized, the resulting NQS has unit norm $\braket{\psi_{\vb*{\theta}}} = 1$~\cite{Sharir2020}.
 
The conditional structure also enables one to perform direct sampling of $p(\vb*{\sigma}) =|\psi_{\vb*{\theta}}(\vb*{\sigma})|^2$. This is a key advantage of autoregressive ansätze with respect to more traditional ones, which rely on Markov-chain Monte Carlo sampling, since here the obtained samples will be uncorrelated. Recurrent neural-network architectures, commonly used in the machine learning literature, such as the gated recurrent unit (GRU) and long-short term memories (LSTM) are inherently autoregressive. For a GRU ansatz, the conditional amplitudes are given by
\begin{equation}
    \phi_i(\sigma_i|\sigma_{i-1} \ldots \sigma_1) = \phi_i(\sigma_i, \vb*{h}_{i-1}) = \text{GRU}(\sigma_i, \vb*{h}_{i-1}),
\end{equation}
where the information on the previous spin variables is contained in the \textit{hidden vectors} $\vb*{h}_i$. The explicit form of the map corresponding to the GRU unit is detailed in Appendix~\ref{app:a}. The number of parameters of a GRU ansatz is fully specified by the dimension of the hidden vector $\vb*{h}_i$, that we denote by $d_h$ in the following, and scales quadratically with $d_h$. The architecture of such an ansatz is schematically represented in Fig.~\ref{fig:scheme_and_SR}(a) for a two-dimensional lattice system.

\subsection{Issues with t-VMC}\label{sec:SR}
As seen previously, in order to propagate a NQS in time according to some Hamiltonian $\hat{H}$, one should minimize the following variational distance:
\begin{align}\label{eq:dist1}
    D(\thetat) = \mathrm{dist}\left(\ket{\psi_{\thetat}}, \exp{-i\hat{H}\delta t}\ket{\psi_{\vb*{\theta}}}\right) %(\hat{\mathds{1}}-i\hat{H}\delta t)\ket{\psi_\theta} \right)
\end{align}
over the variational parameters $\thetat$ after each time step $\delta t$. By Taylor expanding $\ket{\psi_{\thetat}}$ with respect to the variational parameters and keeping only second-order terms, one obtains the following dynamical equation ~\cite{Carleo_2017}:
\begin{equation}\label{eq:SR}
	    S_{k,k'}\dot{\theta}_{k'} = -iF_k,
\end{equation}
with $\dot{\theta}_k := \partial_t \theta_k$, $F_k =  \langle O^*_k E_{\mathrm{loc}}\rangle - \langle O^*_k\rangle \langle E_{\mathrm{loc}}\rangle$, and 
$S_{k,k'} = \langle O^*_k O_{k'}\rangle - \langle O^*_k\rangle \langle O_{k'}\rangle$, known as the quantum geometric tensor~\cite{Stokes_2020}, where the previous expectation values are calculated on the state $\ket{\psi_{\vb*{\theta}}}$. The $O_k$ terms are the log-derivatives of the NQS ansatz, as given by
\begin{equation}
    O_k(\vb*{\sigma}) = \partial_{\theta_k}\ln\psi_{\vb*{\theta}}(\vb*{\sigma}).
\end{equation}
The local energy $E_{\mathrm{loc}}$ is defined as
\begin{equation}
    E_{\mathrm{loc}}(\vb*{\sigma}) = \sum_{\vb*{\sigma}'} \frac{\psi_\theta(\vb*{\sigma}')}{\psi_\theta(\vb*{\sigma})} \mel{\vb*{\sigma}}{\hat{H}}{\vb*{\sigma}'}.
\end{equation}
    The parameters $\vb*{\theta}$ are then updated at each time step according to their derivatives $\dot{\theta}_k$ using a numerical solver. This procedure is known as t-VMC~\cite{becca_sorella_2017, Carleo_2017} and is closely related to natural gradient descent in machine learning~\cite{Goodfellow_2016}. Several difficulties can arise. Firstly, the $\vb{S}$ matrix to be inverted is in general singular, which makes the process extremely sensitive to stochastic variations coming from sampling. Hence, regularization must be used to obtain a non-diverging derivative $\dot{\theta}_k$ \cite{hofmann2021role} of the parameter vector. This increases the stability of the method, although impacting its accuracy, making it challenging to obtain accurate long-time dynamics. Secondly, state-of-the-art regularization techniques involve the inversion of the $\vb{S}$ matrix via its singular-value decomposition, which makes the complexity of the method of order $O(N_{\text{par}}^3)$, with $N_{\text{par}}$ the number of parameters for the NQS.\footnote{Alternatively, one can also perform the inversion with iterative methods such as conjugate gradients, thereby reducing the complexity. This, however, does not enable all forms of regularization.} This is one of the reasons why natural gradient and second-order optimization protocols is rarely used for other machine-learning applications, involving models with up to billions of parameters. Thirdly, for a NQS ansatz with real parameters, one may split the real and imaginary parts of Eq.~\eqref{eq:SR} and solve either of the two resulting equations~\cite{Yuan2019}:
\begin{align*}
    \mathrm{Re}(S_{kk'})\dot{\theta}_{k'} &= 0, \\
     \mathrm{Im}(S_{kk'})\dot{\theta}_{k'} &= - F_k. 
\end{align*}
Solving the first forces one to impose a condition on $\dot{\theta}_k$ to obtain a nontrivial solution ($\dot{\vb*{\theta}}\neq \vb*{0}$), while solving the second one is difficult in general since the diagonal elements of $\mathrm{Im}(\vb{S})$ are zero ($\vb{S}$ is Hermitian) and remaining off-diagonal elements are close to $0$ for various ansätze with real parameters we have tested. This poor conditioning makes regularization schemes even harder and yields large parameter time derivatives, thereby requiring smaller time steps.

To see explicitly such issues with t-VMC, let us consider for example the transverse field Ising Hamiltonian:
\begin{gather}\label{eq:TFI_ham}
   \hat{H}_{\mathrm{TFI}} = -J\sum_{\langle m,n\rangle}  \ssz_m\ssz_{n} + g\sum_m \ssx_m,
\end{gather}
with $J$ the nearest-neighbor coupling strength, and $g$ the transverse field strength. At zero temperature, this model exhibits a quantum phase transition for $g_c = J$ in one dimension and for $g_c \simeq 3.044J$ in two dimensions~\cite{Blote_2002}. For $J>0$, the transition separates a ferromagnetic~\footnote{or anti-ferromagnetic if $J<0$. For simplicity we will consider $J>0$ throughout the paper.} phase from a paramagnetic phase, where the spins tend to align with the transverse-field. In the former, when $g\ll J$, the ground state is degenerate and belongs to the space spanned by $\ket{\uparrow, \uparrow, \ldots ,\uparrow}$ and $\ket{\downarrow, \downarrow, \ldots \downarrow}$, while in the latter, when $g \gg J$ the ground state is $\ket{\rightarrow, \rightarrow, \ldots ,\rightarrow}$, with $\ket{\rightarrow}$ the eigenstate of $\ssx$. This model serves as a convenient benchmark for NQS dynamics ~\cite{Carleo_2017, Schmitt_2020, gutierrez2021real} and is of high interest experimentally, as it has been successfully implemented on quantum simulators~\cite{Schauss_2018}.

In Fig.~\ref{fig:scheme_and_SR}(b), the spectra of $\vb{S}$, computed on the ground state of the TFI Hamiltonian, is shown for three different ansätze: a restricted Boltzmann machine (RBM), an autoregressive convolutional network with complex parameters (ARNN), and the GRU ansatz presented in the previous section, for different values of $g/g_c$ (we have fixed $J=1$) for a two-dimensional array of $N=16$ spins. Two key observations can be made from this figure: firstly, the range of vanishing eigenvalues grows as $g/J$ decreases, something that makes t-VMC less applicable to ferromagnetic-like states. This is the case for both for ground-state search and dynamics, which involve the full spectrum of the Hamiltonian. One can also see that the RBM spectrum is the less singular, and presents a shell-like structure, making it simpler to regularize. In contrast, both the convolutional autoregressive network and GRU ansätze's spectra indicate that the variational space is extremely flat for all values of $g/J$, as very few eigenvalues are of a high enough value, and the magnitude of the eigenvalues decrease extremely rapidly. This inherent poor conditioning implies that if such states are the initial states of a dynamical evolution, the inversion of $\vb{S}$ will yield a parameter update $\dot{\vb*{\theta}}$ whose norm is large with respect to $\lVert \vb*{\theta} \rVert$, forcing one to dramatically reduce the time step which makes calculations impractical. This last spectrum is tricky to regularize, as there is no clear separation of the eigenvalues.

\section{Variational Runge-Kutta algorithms}
As identified in the previous section, t-VMC suffers from a number of limitations that make it unsuitable for use with certain variational ansätze, in particular for the GRU ansatz. We propose to circumvent these issues by resorting back to the original Dirac-Frenkel variational principle (Eq.~\eqref{eq:dist1}) and solving an optimization problem at each time step. The general form of this problem is the minimization of the distance
\begin{align}\label{eq:dist2}
    D(\tilde{\vb*{\theta}}) = \mathrm{dist}\Bigl(\ket{\psi_{\tilde{\vb*{\theta}}}}, \hat{T}\ket{\psi_{\vb*{\theta}}} \Bigr),
\end{align}
where $\tilde{\vb*{\theta}}$ denotes the set of variational parameters to optimize, $\vb*{\theta}$ those at the previous time step $t$, and $\hat{T}$ a propagator evolving the state of the system between times $t$ and $t + \delta t$ under the action of the Hamiltonian of interest. While expanding the propagator to first order in $\delta t$ yields a valid first-order update of the variational state, we instead propose to build a variational principle upon a $s$-order Runge-Kutta approximant, as generated by a propagator $\hat{T}_s$ such that:
\begin{figure*}[t]
    \centering
    \includegraphics[width=\textwidth]{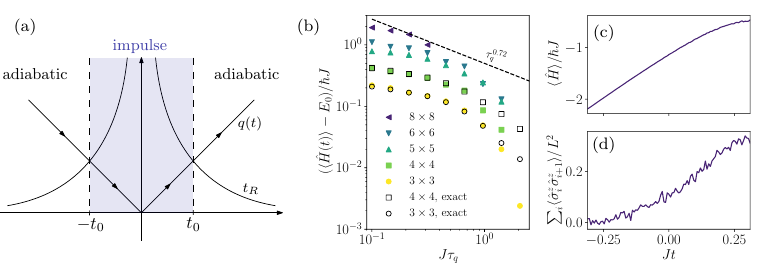}
    \caption{(a) Scheme of the Kibble-Zurek mechanism. At times $\pm t_0$, the relaxation time of the system $t_R(t_0)$ crosses the time scale of the quench $q(t_0)$, which defines the limit between the adiabatic region, where the system essentially stays in the ground state of the system at that time, and the impulse regime, where nonequilibrium dynamics take place. (b) Scaling of the injected energy $\langle \hat{H}(t) \rangle - E_0$ as a function of the quench time $\tau_q$ for different system sizes $N = L\times L$, with $t=\tau_q$ (the final time). One can see that for fast quenches the NQS dynamics predict very well the injected energy. The data for system sizes with more than $L = 5$ cannot be simulated via exact diagionalization, hence they are compared to the $\tau_q^{0.72}$ scaling obtained in Ref.~\cite{Schmitt_2021}. (c) Average energy for a quench of total time $J\tau_q = 0.31$, for a $N = 6\times 6$ system and a GRU ansatz with $d_h = 8$, with $N_s = 200$ samples. (d) Average correlation function along the $z$ axis for the same quench as in (c). 
    }
    \label{fig:2DKZ_scaling}
\end{figure*}
\begin{equation}
    \ket{\psi(t+\delta t)} = \hat{T}_s(t) \ket{\psi(t)} + O(\delta t^{s+1}).
\end{equation}
General explicit expressions for this operator are derived in Appendix~\ref{app:a} for arbitrary diagonally implicit Runge-Kutta schemes. In particular, for the standard (second-order) Heun method used throughout the paper, one has:
\begin{equation}\label{eq:Heun_prop}
    \hat{T}_\mathrm{Heun}(t) = \hat{\mathds{1}} - i\delta t \hat{H}(t) - \frac{\delta t^2}{2} \hat{H}(t+\delta t)\hat{H}(t).
\end{equation}
The variational method induced by this scheme involves an integration error per time step of third order in $\delta t$. In practice, rather than the Fubini-Study distance, we use the following numerically well-behaved metric based upon the quantum fidelity:

\begin{equation}
    \mathrm{dist}\Bigl(\ket{\psi}, \ket{\phi}\Bigr) = 1 - \frac{\lvert\braket{\psi}{\phi}\rvert^2}{\braket{\psi}\braket{\phi}},
\end{equation}
which for normalized quantum states reduces to
\begin{align}\label{eq:dist}
    D(\tilde{\vb*{\theta}}) = 1 - \frac{|\mel{\psi_{\tilde{\vb*{\theta}}}}{\hat{T}_s}{\psi_{\vb*{\theta}}}|^2}{\mel{\psi_{\tilde{\vb*{\theta}}}}{\hat{T}_s^\dagger\hat{T}_s}{\psi_{\vb*{\theta}}}}.
\end{align}
For the Heun method, this may be simplified as
\begin{equation}\label{eq:simp_dist}
    D(\tilde{\vb*{\theta}}) = 1 - |\mel{\psi_{\tilde{\vb*{\theta}}}}{\hat{T}_{\mathrm{Heun}}}{\psi_{\vb*{\theta}}}|^2 + O(\delta t^{4}),
\end{equation}
with
\begin{equation}
    \mel{\psi_{\tilde{\vb*{\theta}}}}{\hat{T}_s}{\psi_{\vb*{\theta}}} \equiv \mel{\psi_{\tilde{\vb*{\theta}}}}{\hat{T}_\mathrm{loc}}{\psi_{\tilde{\vb*{\theta}}}},
\end{equation}
where $\hat{T}_\mathrm{loc}$ is an operator acting as a local estimator and whose non-zero entries are given by
\begin{align}\label{eq:t_loc}
T_{\mathrm{loc}}(\vb*{\sigma}) :=& \mel{\vb*{\sigma}}{\hat{T}_\mathrm{loc}}{\vb*{\sigma}}\nonumber\\
    =& \sum_{\vb*{\sigma}'} \frac{\psi_{\vb*{\theta}}(\vb*{\sigma}')}{\psi_{\tilde{\vb*{\theta}}}(\vb*{\sigma})} \mel{\vb*{\sigma}}{\hat{T}_s}{\vb*{\sigma}'}.
\end{align}
Technical details on the derivation of Eq.~\eqref{eq:simp_dist} are provided in Appendix~\ref{app:e}.

It thus appears from the above that any $s$-order Runge-Kutta update can be cast into a variational problem consisting in maximizing the squared expectation value of an observable. Furthermore, this expectation value can be efficiently sampled at every step of the optimization process as the average
\begin{align}
    \mel{\psi_{\tilde{\vb*{\theta}}}}{\hat{T}_\mathrm{loc}}{\psi_{\tilde{\vb*{\theta}}}} = \mathbb{E}_{\boldsymbol{\sigma}\sim|\psi_{\tilde{\boldsymbol{\theta}}}|^2}[T_\mathrm{loc}(\vb*{\sigma})].
\end{align}
over configurations $\vb*{\sigma}$ drawn from the probability distribution corresponding to $\ket{\psi_{\tilde{\vb*{\theta}}}}$. Note that this involves summing over the connected elements of $\hat{T}$ as can be seen in Eq.~\eqref{eq:t_loc}. This will in general become more expensive as the order $s$ increases, as powers of the Hamiltonian will be contained in $\hat{T}$. The number of connected elements will depend on the sparsity of the Hamiltonian; for the transverse-field Ising model for instance, the number of connected elements in $\hat{H}^s$ scales as $N^{s}$ (which means $N^2$ for a second-order integration scheme). \kd{Therefore, the computational cost of the presented method scales as $O(N^s n_s N_{\text{par}}n_g)$ per time step, with $n_s$ the number of samples, $N_{\text{par}}$ the number of variational parameters and $n_g$ the number of gradient evaluations, or minimization steps. This comes from the minimization of the distance that includes $n_g$ gradient calculations at each time step. This scaling can be compared to the cost of the inversion of the $\mathbf{S}$ matrix in t-VMC that scales as $O(N_{\text{par}}^3)$ for regularization techniques to be used. We therefore believe our method is better suited for large neural network architectures, as it only scales linearly with the number of parameters.}
\begin{figure*}
    \centering
    \includegraphics[scale=0.7]{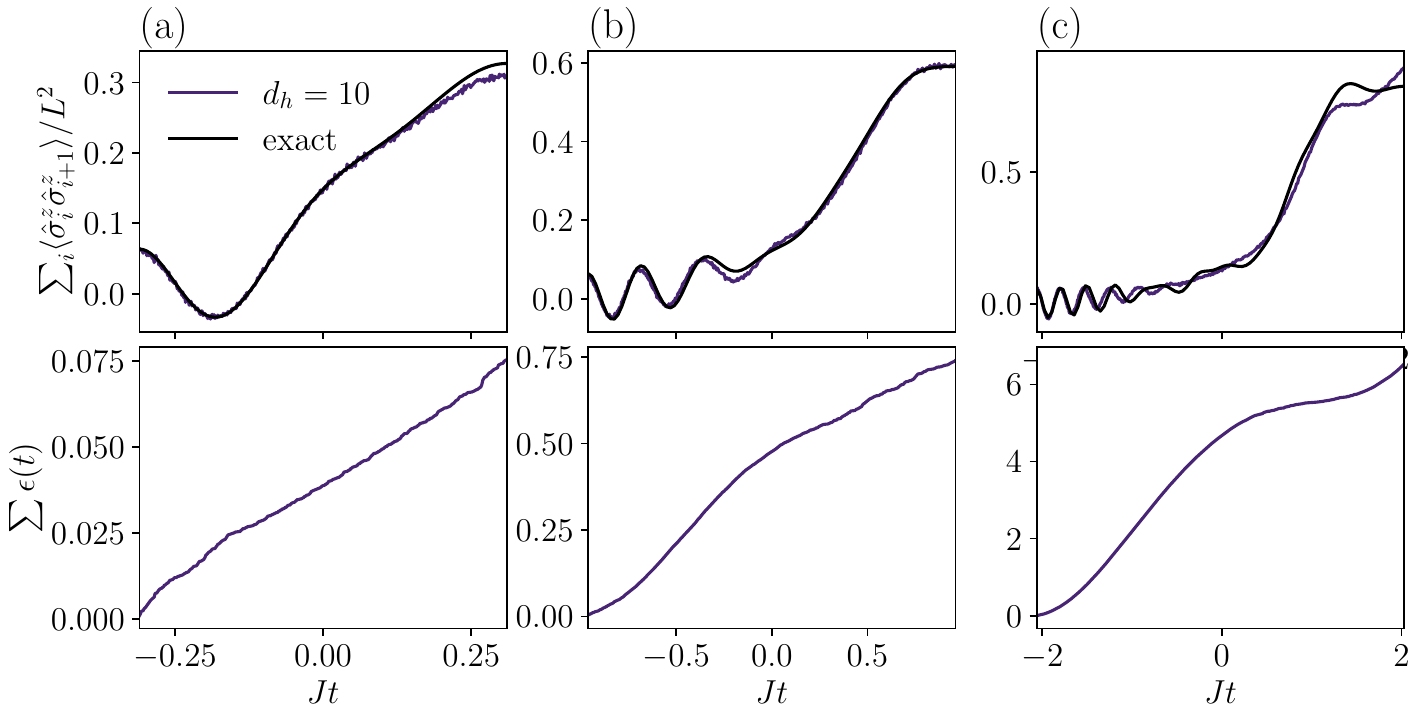}
    \caption{Numerical results obtained for a linear quench for various values of the total quench time $\tau_q$. On the upper panels, the average correlation function is shown as a function of time for both the exact (dashed line) and the GRU simulations (solid lines). The associated cumulative error as a function of time is shown in the lower panels (note the scales of the vertical axis). For panels (b) and (c), the time step was chosen to be linearly decreasing, as the characteristic time scale changes as a function of time. Parameters for each quench are given in Appendix~\ref{app:B}. Here $N = 4 \times 4$ spins.}
    \label{fig:benchmark_KZ}
\end{figure*}

\section{Application: critical quenches}

\subsection{The Kibble-Zurek mechanism}

The Kibble-Zurek mechanism~\cite{Kibble_1976, zurek1985} predicts the formation of topological defects in quenched systems undergoing a second-order phase transition as the system parameters are linearly swept across a critical point. This universal behavior stems from the fact that there exists a time, denoted $t_0$, at which the characteristic evolution time of the system (the relaxation time $t_R(t_0)$, related to the inverse of the energy gap) becomes larger than the characteristic quench timescale $q(t_0) \sim |t_0/\tau_q|$, as schematically shown in Fig.~\ref{fig:2DKZ_scaling}(a). Before $t_0$, the dynamics remains quasi-adiabatic, as the Hamiltonian parameters are tuned slowly with respect to the characteristic time scale of the system. However, after this time, genuinely non-equilibrium dynamics takes place as critical slowing down sets in and the system dynamics becomes slower than the Hamiltonian parameter sweep. Ignoring specific details of this dynamics, one can derive a scaling law for the density of created defects at the end of a linear quench~\cite{zurek1985}, namely
\begin{align}
    \langle \hat{n}_d(\tau_q) \rangle \sim \tau_q^{-d\nu/(z\nu + 1)}
\end{align}
with $d$ the dimensionality of the system and $z,\nu $ universal critical exponents. Note that $\hat{n}_d$ will depend on the geometry of the system, and is not always easily accessible experimentally. For this reason some works consider other observables, such as the number of domain walls~\cite{Puebla_2019}, or the injected energy density~\cite{Schmitt_2021}, defined as
\begin{align}
    \mathcal{E} = \frac{1}{L^2} \left[\langle \hat{H}(t) \rangle - E_0(t) \right],
\end{align}
where $E_0$ denotes the ground-state energy, which is a witness of the injected defects. These are excitations of the system with respect to its ground state at time $t$, hence $\mathcal{E}$ must scale as the density of defects. This quantity is convenient as it does not depend on the geometry of the physical system under consideration. In Ref.~\cite{Schmitt_2021}, authors extract a scaling law going as $\tau_q^{0.72}$ for the 2D transverse-field Ising model based upon various numerical methods that they used to extract a correlation length by estimating the gap of the system for different values of $g/g_c$. We consider a similar protocol in this paper, in which both $J$ and $g$ are varied in time. The Hamiltonian reads
\begin{gather}
   \hat{H}_{\mathrm{TFI}}/\hbar = -J(t)\sum_{\langle m,n\rangle}  \ssz_m\ssz_{n} + g(t)\sum_m \ssx_m,
\end{gather}
 \begin{figure*}
    \centering
    \includegraphics[scale=0.6]{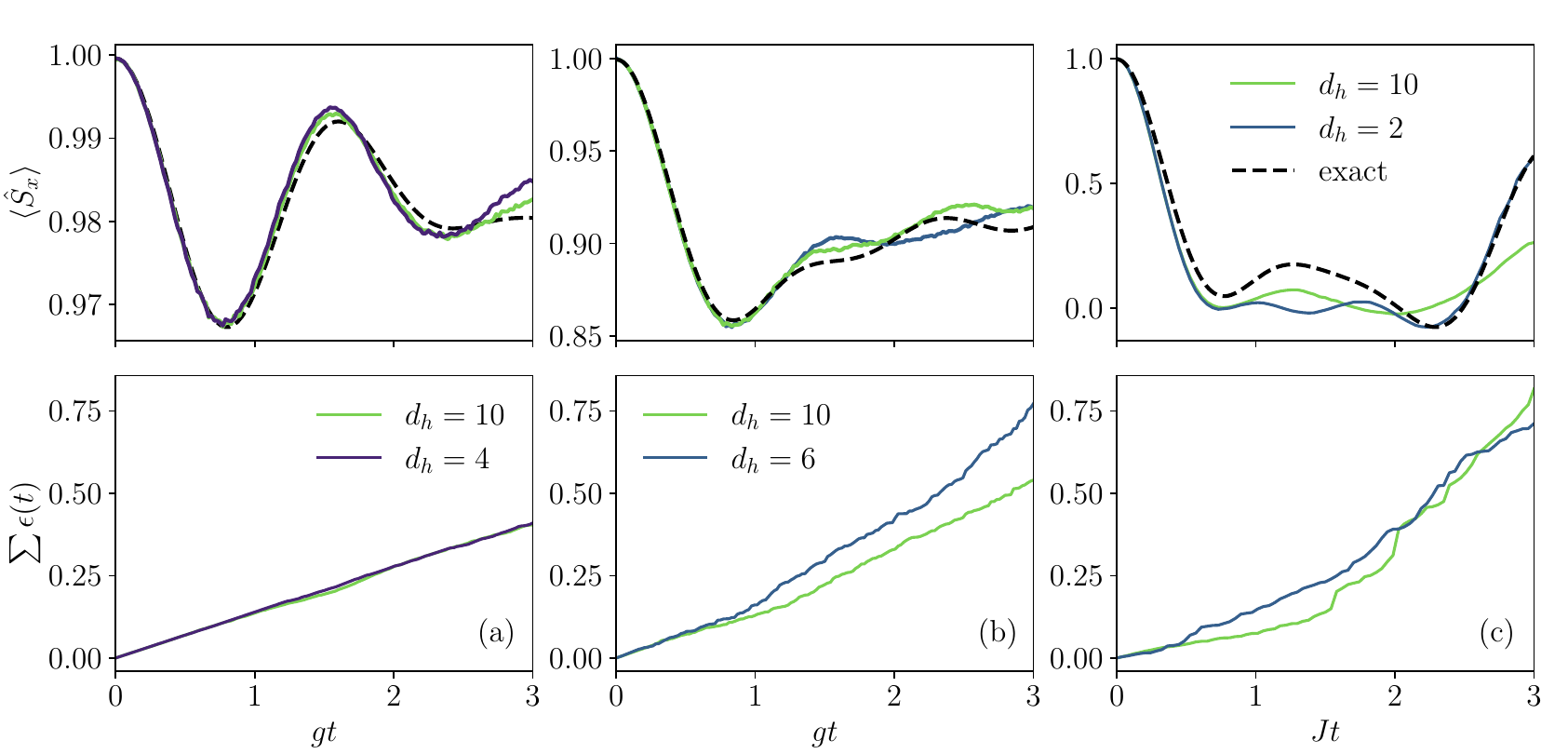}
    \caption{Dynamics induced by suddenly quenching the Hamiltonian parameters to $g/g_c = 2$, $g/g_c = 1$ and $g/g_c=1/10$ for various hidden-vector dimensions $d_h$, starting from an initial paramagnetic state. Upper panels show the average magnetization $\langle \hat{S}_x\rangle = \sum_i \hat{\sigma}^x_i / L^2$ and the lower panels show the cumulative error as a function of time. Here $N = 4 \times 4$ spins.}
    \label{fig:benchmarks_IQ}
\end{figure*}
with $J(t) = J(1+t/\tau_q)$, $g(t) = g_0(1-t/\tau_q)$ and $g_0 = g_c$. This type of quench involves all regimes ($0 \leq g/J < +\infty$), hence t-VMC does not enable one to access the full dynamics in two dimensions~\cite{Schmitt_2021}. However, our approach does not suffer from such issues, which is one of its major strengths. In Fig.~\ref{fig:2DKZ_scaling}(b), we show the injected energy (not rescaled for readability) $\langle \hat{H}(t) \rangle - E_0(t)$ for various system sizes, for different values of the quench time. One can see that the exact results are closely matched for fast quenches, and precision drops for slower quenches ($J\tau_q > 1$). This is related to the fact that dynamics become almost fully adiabatic, hence the NQS dynamics is not precise enough to capture the small amount of injected energy. The presented method most accurately reproduces the dynamics of the system in the non-adiabatic region, where the dynamics presents universal features; this is the regime of relevance when studying the Kibble-Zurek mechanism. For lattices above $5\times 5$, we can no longer compare our results with exact-diagonalization, but the scaling predicted in~\cite{Schmitt_2021} is recovered, indicating that the results are reliable. \kd{We also note that for small systems, such as the $3\times 3$ system, inaccuracies appear for large values of $J\tau_q$, which corresponds to slow quenches. This is due to the system remaining in an adiabatic regime, since its gap is larger. Hence the value of the injected energy decreases, requiring more samples and a smaller time step to be resolved within a comparable relative error. However, when investigating Kibble-Zurek quenches, one is mainly interested in the (universal) non-adiabatic regime.}
In Fig.~\ref{fig:benchmark_KZ}, we also show correlation functions for various quench times, as well as the residual error corresponding to the full dynamics, as given by
\begin{align}
    \epsilon(t) &= \mathrm{dist}\Bigl(\ket{\psi_{\tilde{\vb*{\theta}}}}, \hat{T}\ket{\psi_{\vb*{\theta}}} \Bigr)\nonumber\\
     &= 1 - |\langle \psi_{\tilde{\vb*{\theta}}}|\hat{T}\ket{\psi_{\vb*{\theta}}}|^2.
\end{align}
One observes that for panels (a) and (b), corresponding to fast quenches, the dynamics is accurately captured, while for panel (c), corresponding to a slower quench, the dynamics is reproduced although with poorer accuracy as $t$ approaches $\tau_q$. Note also that the cumulative error $\sum_t \epsilon (t)$ increases by an order of magnitude from panel to panel. This can be partially ascribed to the larger number of time steps required to faithfully simulate the dynamics of longer quenches.

\subsection{Sudden quenches}
 One can also investigate the nonequilibrium dynamics of many-body systems without relying on regularization hyperparameters and instabilities stemming from t-VMC. 
 As a benchmark of the presented method, we consider the two-dimensional time-independent transverse-field Ising model, defined in Eq.~\eqref{eq:TFI_ham}.
\begin{figure*}
    \centering
    \includegraphics[scale=0.5]{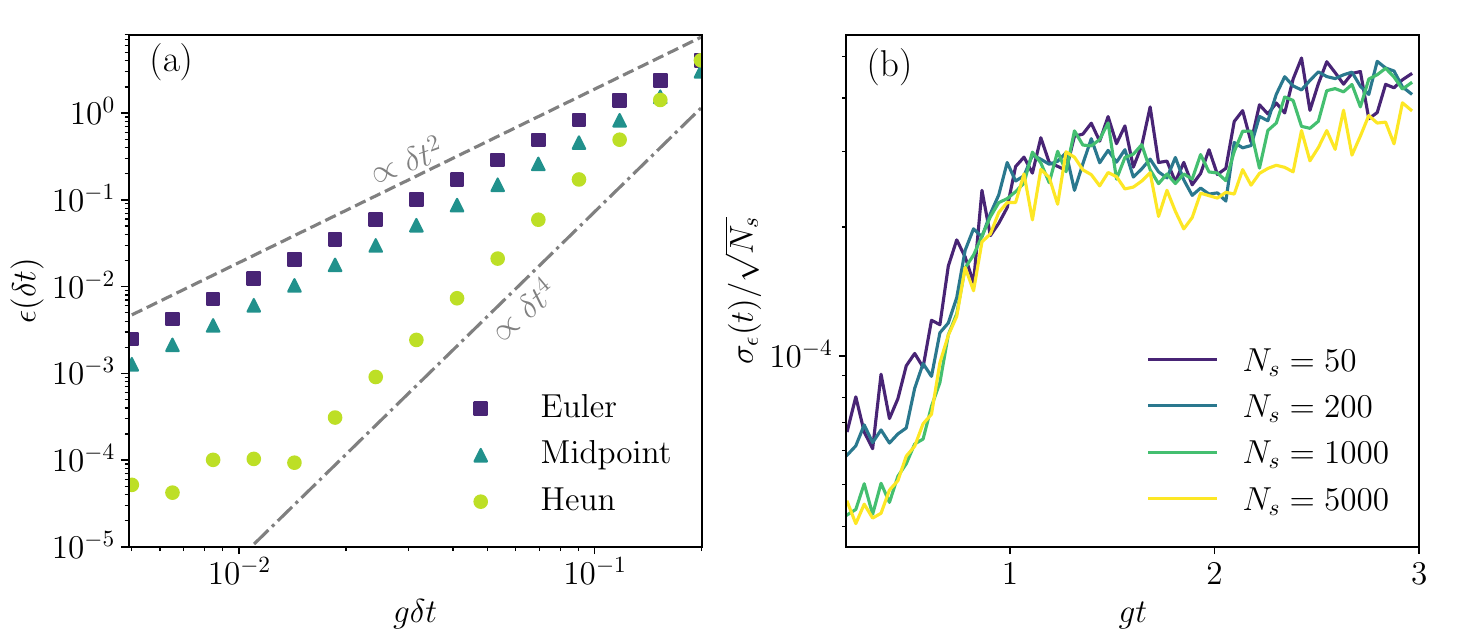}
    \caption{(a) Residual error $\epsilon$ as a function of the time step $\delta t$ for the Euler method, the midpoint method, and the Heun integration methods averaged over the 10 first time steps of the quench dynamics for the transverse-field Ising model with $g/g_c = 2$. (b) Standard deviation of the error as a function of time rescaled by $\sqrt{N_s}$. Curves collapse, which is what is expected from direct sampling. This enables us to estimate the required number of samples to obtain a given error. Note that as the dynamics progresses, the variance of the residual error increases, indicating a increasing difficulty in reaching the optimum in time. Here $N = 4 \times 4$ spins.}
    \label{fig:error}
\end{figure*}
We prepare the system in the ground state of the TFI Hamiltonian for $g \gg J$, which is $\ket{\psi_0} = \ket{\rightarrow, \rightarrow, \ldots ,\rightarrow}$, and quench the magnetic field to the values $g/g_c= 2$, $1$, and $1/10$. The corresponding results are displayed in panels (a), (b), and (c) of Fig.~\ref{fig:benchmarks_IQ} respectively. By doing this, one spontaneously creates excitations of all the eigenstates of the Hamiltonian corresponding to the final value of the parameters. This is of high interest experimentally, as it can be used to probe the properties of non-integrable systems close to the critical point. It is expected that a quench near the critical point is the most difficult to simulate, since it involves states that are correlated at all scales in the thermodynamical limit. In contrast to other approaches, the use of direct sampling here ensures that the dynamics will be accurate for a lower number of samples. Here, we have imposed the reflection  and $\mathds{Z}_2$ symmetries for panels (a) and (b), and no symmetry for panel (c). Surprisingly, in this last case, imposing symmetries, leads to a worsened accuracy as we show in Appendix~\ref{app:C}. 

\subsection{Error analysis}

The residual error $\epsilon$ depends on the chosen time step $\delta t$. In Fig.~\ref{fig:error}, this is shown as a function of the chosen time step $\delta t$ for the Euler, (implicit) midpoint, and Heun integration schemes. The propagators $\hat{T}$ for the Euler and the implicit midpoint method are respectively given by
\begin{align}
    \hat{T}_\mathrm{Eul} &=  \hat{\mathds{1}} - i\delta t \hat{H},\\
    \hat{T}_\mathrm{mid} &= \hat{\mathds{1}} - i\delta t \hat{H} +\frac{\delta t^2}{4} \hat{H}^2.
\end{align}
The Heun propagator is given by Eq.~\eqref{eq:Heun_prop}. Euler is a first-order method, while the midpoint and Heun methods are both second-order methods. The midpoint method conserves a symplectic symmetry in the exact case, which for the Schrödinger equation corresponds to energy conservation. Note, however, that by variationally propagating an NQS in time, this is no longer guaranteed, as the time evolution is approximated stochastically. Using $\hat{T}_\mathrm{mid}$ as a propagator is equivalent to minimizing the distance proposed in Ref.~\cite{gutierrez2021real} and is expected to yield a second-order update. However, an integration error of order $O(\delta t^2)$, characteristic of first-order methods, is instead observed in Fig.~\ref{fig:error} for this scheme. This poorer scaling can be ascribed to $\hat{T}_\mathrm{mid}$ being non-unitary to second order in $\delta t$. Interestingly, as shown in Appendix~\ref{app:e}, this method requires the norm of the NQS to depart from unity to second order in $\delta t$ to match its optimal order of accuracy, making it effectively first-order when applied to normalized ansätze. 
In contrast, the $\delta t^4$ scaling of the error of the Heun method is found to be better than the usual $\delta t^3$ scaling of second-order methods.
Hence, in this context, the midpoint method carries the disadvantage of having an error comparable to an Euler update while having the computational complexity of the Heun method. One can also see that for the Heun method there is a threshold around $g\delta t = 0.01$ past which the error cannot be reduced by decreasing the time step. This is due to the systematic error stemming from the optimization process for a given set of hyperparameters. Nevertheless, this minimal error bound can be decreased by tuning the hyperparameters, and, in particular, by increasing the number of samples. Note that in the constant-error regime, reducing the time step is detrimental as, for a given fixed total simulation time, a larger number of time steps is required, thereby proportionally increasing the total error. There is therefore an optimal time step, which we find to be $g\delta t \approx 0.01$ for the various hyperparameters used in this work. In Fig.~\ref{fig:error}(b), the standard deviation of the error is shown for each time step of a sudden quench from the paramagnetic phase to $g/g_c = 2$. One can distinguish two regimes: the standard deviation of the residual error first grows till $gt \approx 1$, and then saturates. This can be attributed to a harder optimization after a given time, where an optimum is indeed found but with a greater variance as the quench goes on. These results are reminiscent of those recently reported in Ref.~\cite{lin2021scaling}, where authors demonstrate that quantum states become harder to fit with an NQS at later times after a sudden quench, which is not always related to the growth of entanglement. Standard deviations of the error are shown for increasing  values of $N_s$, the number of samples considered for each optimization step. One can see that $\sigma_\epsilon(t)$ indeed decreases as $\sqrt{N_s}$, as expected from direct sampling. 

\section{Conclusion}
We have presented an alternative scheme for the real-time evolution of quantum many-body systems with an NQS ansatz. This scheme does not rely on hyperparameter tuning for regularization and is found to be stable, and is not limited by the regime of the physical system. We therefore expect it to be useful when t-vMC fails, when one uses autoregressive models or when the network contains a large number of variational parameters. This scheme could be used to further investigate the limits of the Kibble-Zurek mechanism in nontrivial quantum systems, such as frustrated systems in triangular lattices or in systems with multiple critical points. As other applications, quantum control~\cite{Dong_2010} and pulse optimization problems~\cite{Li_2022} could also be considered, where accurate methods to simulate time-dependent dynamics are required. As an outlook, the scheme could further be improved by considering different cost functions that lead to weaker optimization errors and variance. We also expect the scheme to be helpful for the dynamics of open quantum systems~\cite{Hartmann_2019, Reh_2021, Vicentini_2022} or to compute individual quantum trajectories in approaches such as Monte Carlo wavefunction~\cite{Carmichael_1993,Molmer_1996,Carmichael_2008} or corner methods based on low-rank representations of the density matrix~\cite{Donatella_2021}.

\begin{acknowledgments}
		We would like to acknowledge stimulating discussions with F. Vicentini. This work was supported by ANR, via the projects UNIQ (ANR-16-CE24-0029) and TRIANGLE (ANR-20-CE47-0011), and by the FET FLAGSHIP Project PhoQuS (grant agreement ID: 820392). We also acknowledge access to the high performance computation center TGCC of the French national computational facility GENCI under the projects 2021-A0100512462 and 2022-A0100512462. 
	\end{acknowledgments}

\bibliography{bib}
\clearpage
\onecolumngrid
\appendix

\section{Models\label{app:a}}

Here we present the different variational ansätze we considered in the paper. In section~\ref{sec:SR}, we compare the spectra of the quantum geometric tensor $\vb{S}$ for three different ansätze.

\noindent\textbf{Restricted Boltzmann machine (RBM)} -- The first ansatz we used is the RBM, whose amplitudes are defined as:
\begin{equation}
    \Psi_{\params}(\vb*{\sigma}) = e^{\biases_v^\dagger \vb*{\sigma}}\prod_{i=1}^{M}2\cosh\left(\biases_{h,i} + \vb{W}_{i,:}\vb*{\sigma}\right),
    \label{eq:NQS_rbm_amp}
\end{equation}
where $\biases_{h,i}$ and $\vb{W}_{i,:}$ denote the $i$th hidden bias and weight matrix row, respectively, and $\biases_v$ denotes the visible bias.

\noindent\textbf{Convolutional autoregressive neural network} -- This network is a convolutional feed-forward neural network whose convolutional filters have been masked so as to respect the autoregressive property: a directionality of the connections in the neural network is imposed, as the output of the $i$th unit in any layer of the network must ultimately depend only on the local spin configurations $\sigma_{j<i}$. This ansatz roughly resembles a simplified version of the ansatz considered in \cite{Sharir2020}.

\noindent\textbf{Gated recurrent unit (GRU)} -- The third ansatz is the one used throughout the paper, the GRU ansatz, a variant of the recurrent neural network. The conditional amplitudes for this ansatz are obtained through the following transformation, which depends on $\sigma_i$, the local spin, and $\vb*{h}_{i-1}$, the hidden unit coming out of the previous RNN cell:
\begin{align}
    \phi_i(\sigma_i, \vb*{h}_{i-1}) := \text{GRU}(\sigma_i, \vb*{h}_{i-1}) &= A(\sigma_i)\exp{i\varphi(\sigma_i)},\\
    A(\sigma_i) &= \varsigma(\vb*{U}_A \vb*{h}_i + \vb*{b}_A), \\
    \varphi(\sigma_i) &= \varsigma\varsigma(\vb*{U}_\varphi \vb*{h}_i + \vb*{b}_\varphi),
\end{align}
where $\varsigma$ and $\varsigma \varsigma$ denote a softmax and softsign transformation respectively, and where we have defined the conditional modulus $A(\sigma_i)$ and phase $\varphi(\sigma_i)$ that depend on matrices $\vb*{U}_A$ and $\vb*{U}_\varphi$ respectively, which each contain variational parameters. The hidden vector $\vb*{h}_i$ is given by
\begin{align}
    \vb*{h}_i = (1 - \vb*{z}_j) \odot \vb*{h}_{i-1} + \vb*{z}_j \odot \tilde{\vb*{h}}_j
\end{align}
where $\odot$ denotes the Hadamard product and the \textit{latent hidden vector} $\tilde{\vb*{h}}_j$ is given by
\begin{align}
   \tilde{\vb*{h}}_j = \tanh{\left(\tilde{\vb{W}}[\vb*{r}_j \odot \vb*{h}_{j-1},\vb*{\sigma}_j] + \tilde{\vb*{b}} \right)} 
\end{align}
and the \textit{update gate} $\vb*{z}_j$ and \textit{reset gate} $\vb*{r}_j$:
\begin{align}
   \vb*{z}_j &= \mathrm{sig}\left(\vb*{W}_z[\vb*{h}_{j-1} ; \vb*{\sigma}_{j-1}] - \vb*{b}_z \right),\\
    \vb*{r}_j &=\mathrm{sig}\left(\vb*{W}_r[\vb*{h}_{j-1} ; \vb*{\sigma}_{j-1}] - \vb*{b}_r \right),
\end{align}
where sig denotes the sigmoid funtion and where we have defined the matrices $\tilde{\vb*{W}}, \vb*{W}_r, \vb*{W}_z$ and bias vectors $\tilde{\vb*{b}}, \vb*{b}_r, \vb*{b}_z$ which are all variational parameters. $[\vb*{h}_{j-1} ; \vb*{\sigma}_{j-1}]$ denotes a concatenation of vectors $\vb*{h}_{j-1}$ and $\vb*{\sigma}_{j-1}$, the latter corresponding to the one-hot encoding of the local spin configuration $\sigma_{j-1}$. The important point here is that the total number of variational parameters scales quadratically with the dimension of the hidden vector $d_h$, since the $\vb*{W}$ matrices each contain $d_h\times (d_h+2)$ variational parameters for a one-hot encoding of the local spins $\sigma_i$.
One can see from this transformation that the GRU ansatz naturally respects the autoregressive property.

\section{Computational details\label{app:B}}
The exact simulations were performed with QuTiP~\cite{Johansson_2012}, and the variational simulations and optimization were performed thanks to NetKet~3 modules~\cite{netket_2021}.
In the following table we show the parameters used to produce the results in the main text. For all the simulations we used the Adam optimizer with $b_1 =0.9, b_2 = 0.999, \epsilon=10^{-8}$.\\
\begin{center}
     \begin{tabular}{c|c|c}
   Figure & Network details & Parameters\\
   \hline
Fig. 1(b) & RBM, complex parameters, $\alpha = 2$&  $n_{\mathrm{steps}} = 500, \eta = 0.01, N_s = 1000$ \\
Fig. 1(c)    & convARNN, complex parameters, 3 layers  &  $n_{\mathrm{steps}} = 500, \eta = 0.01, N_s = 1000$\\
Fig. 1(d)  & GRU $d_h = 10$ &  $n_{\mathrm{steps}} = 500, \eta = 0.01, N_s = 1000$\\
Fig. 2, $8\times 8$ & GRU, $d_h = 8$ &$n_{\mathrm{steps}} = 50, \eta = 0.006, N_s = 200$\\
Fig. 2, $6\times 6$ & GRU, $d_h = 8$ &$n_{\mathrm{steps}} = 100, \eta = 0.005, N_s = 200$\\
Fig. 2, $5\times 5$ & GRU, $d_h = 8$ &$n_{\mathrm{steps}} = 100, \eta = 0.01, N_s = 200$\\
Fig. 2, $4\times 4$ & GRU, $d_h = 8$ &$n_{\mathrm{steps}} = 200, \eta = 0.01, N_s = 500$\\
Fig. 2, $3\times 3$ & GRU, $d_h = 8$ &$n_{\mathrm{steps}} = 200, \eta = 0.01, N_s = 1000$\\
Fig. 3 & GRU, $d_h = 10$ &$n_{\mathrm{steps}} = 100, \eta = 0.01, N_s = 1000$\\
Fig. 4 & GRU, -- &$n_{\mathrm{steps}} = 100, \eta = 0.01, N_s = 1000$\\
Fig. 5: (a) & GRU, $d_h = 8$ &$n_{\mathrm{steps}} = 100, \eta = 0.01, N_s = 1000$\\
Fig. 5: (a) & GRU, $d_h = 8$ &$n_{\mathrm{steps}} = 100, \eta = 0.01, N_s = 1000$\\
Fig. 5: (b) & GRU, $d_h = 8$ &$n_{\mathrm{steps}} = 100, \eta = 0.01$\\
Fig. 6: (b) & GRU, $d_h = 10$ & $n_{\mathrm{steps}} = 100, \eta = 0.01, Ns = 500$\\
\hline
\end{tabular}
\end{center}
For Fig. 3 (b) and (c), we have considered a linearly decreasing time step, to account for the fact that during a Kibble-Zurek quench timescales change during the dynamics. This leads to a roughly constant error per time step as can be seen in the corresponding lower panels.
\section{Effect of imposing symmetries}\label{app:C}
\begin{figure}
    \centering
    \includegraphics[scale=0.6]{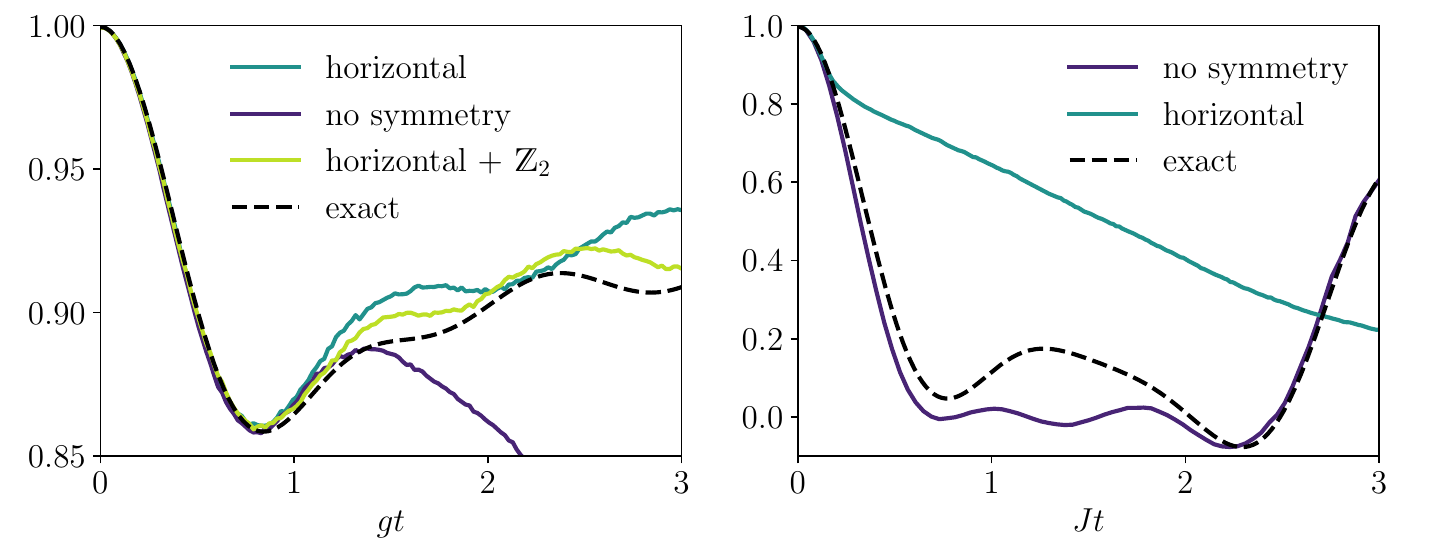}
    \caption{Dynamics of the average magnetization $\langle\hat{S}_x\rangle$ with different symmetries imposed for the quench to the critical point $g = g_c$ (left panel) and to the ferromagnetic phase $g/g_c=0.1$ (right panel) for a $4\times 4 $ lattice.}
    \label{fig:symm}
\end{figure}

In Fig.~\ref{fig:symm} we show the effect of imposing symmetries in two cases, where a sudden quench is performed, both to the critical point and to the ferromagnetic phase. For the critical point, one can see that imposing both $\mathds{Z}_2$ and horizontal permutation symmetries dramatically improves the obtained dynamics. In constrast, imposing any symmetry when performing the dynamics of a ferromagnetic quench, although improving the precision at the beginning, eventually it yields a wrong time evolution.

\section{High-order integration methods for NQS}

\subsection{Generalization to higher-order methods}\label{app:D2}

Let us consider some linear ordinary differential equation of the form
\begin{equation}
	\partial_t \ket{\psi(t)} = \hat{\Phi}_{t}\ket{\psi(t)},
\end{equation}
whose solution $\ket{\psi}$ is discretized over a set of well-defined times $\lbrace t_n\rbrace_{n=0}^{N}$ such that $\ket{\psi^\pidx{n}} \equiv \ket{\psi(t_n)}$. Linear Runge-Kutta (RK) methods provide us with the following prescription for approximately updating $\ket{\psi^\pidx{n}}$:
\begin{align}
	\sket{\psi^\pidx{n{+}1}} &= \sket{\psi^\pidx{n}} + h_n \sum_{i=1}^s b_i \sket{\kappa_i^\pidx{n}},\label{eq:2}\\
	\sket{\kappa_i^\pidx{n}} &= \hat{\Phi}_{t_{n,i}} \bigl(\sket{\psi^\pidx{n}} + h_n{\textstyle\sum_{j=1}^{s}}a_{ij}\sket{\kappa_j^\pidx{n}}\bigr),\label{eq:3}
\end{align}
where $s$ denotes the number of stages of the method, $h_n = t_{n+1} - t_n$, $t_{n,i} = t_n - c_i h_n$ and the coefficients $\lbrace a_{ij}\rbrace_{i,j}$, $\lbrace b_i\rbrace_i$ and $\lbrace c_i\rbrace_i$ are completely determined by the Butcher tableau of the considered method:
\begin{equation}
	\renewcommand\arraystretch{1.2}
	\begin{array}
		{c|cccc}
		c_1 & a_{11} & a_{12} & \cdots & a_{1s}\\
		c_2 & a_{21} & a_{22} & \cdots & a_{2s}\\
		\vdots & \vdots & \vdots & \ddots & \vdots\\
		c_s & a_{s1} & a_{s2} & \cdots & a_{ss}\\
		\hline
		& b_1 & b_2 & \cdots & b_s \, .
	\end{array}
\end{equation}
Note that the entries are nonzero only on the lowest triangular matrix for explicit methods. We will here consider methods whose Butcher coefficients satisfy either $a_{i,j>i} = 0$ (explicit) or $a_{i,j\geq i} = 0$ (implicit).

For the considered tableaus, the system of Eqs.~\eqref{eq:2} and \eqref{eq:3} can be solved explicitly by using the following recurrence identity:
\begin{equation}
	\sket{\kappa_i^\pidx{n}} = \hat{\Pi}_{i}^\pidx{n} \sket{\psi^\pidx{n}} + h_n\sum_{j=1}^{i-1}\hat{\Pi}_{i}^\pidx{n}a_{ij}\sket{\kappa_j^\pidx{n}},
\end{equation}
with
\begin{equation}
	\hat{\Pi}_{i}^\pidx{n} = \bigl[\hat{\mathds{1}} - a_{ii}h_n\hat{\Phi}_{t_{n,i}}\bigr]^{-1}\hat{\Phi}_{t_{n,i}}.
\end{equation}
Indeed, one has:
\begin{equation}
	\sket{\kappa_i^\pidx{n}} = \Bigl\lbrace\hat{\Pi}_i^\pidx{n} + h_n \hat{\Pi}_i^\pidx{n} a^{ij}\hat{\Pi}_j^\pidx{n} + h_n^2 \hat{\Pi}_i^\pidx{n} a^{ij}\hat{\Pi}_j^\pidx{n}a^{jk}\hat{\Pi}_k^\pidx{n} + \ldots\Bigr\rbrace\sket{\psi^\pidx{n}},
\end{equation}
and thus
\begin{equation}
	\sket{\psi^\pidx{n{+}1}} = \hat{T}_s^\pidx{n}\sket{\psi^\pidx{n}},
\end{equation}
with
\begin{equation}
	\hat{T}_s^\pidx{n} = \hat{\mathds{1}} + \sum_{i=1}^s b_i \Bigl(h_n\hat{\Pi}_i^\pidx{n} + h_n^2 \hat{\Pi}_i^\pidx{n} a^{ij}\hat{\Pi}_j^\pidx{n} \\+ h_n^3 \hat{\Pi}_i^\pidx{n} a^{ij}\hat{\Pi}_j^\pidx{n}a^{jk}\hat{\Pi}_k^\pidx{n} + \ldots\Bigr).
\end{equation}

Representing the wavefunctions above with a variational ansatz, namely $\braket{\vb*{\sigma}}{\psi_{\vb*{\eta}_n}} = \psi_{\vb*{\eta}_n}(\vb*{\sigma})$, the generic update can finally be recast into the following optimization process:
\begin{equation}
	\sket{\psi_{\vb*{\eta}_{n+1}}} = \argmin_{\ket{\psi_{\vb*{\eta}}}}\mathrm{dist}\Bigl(\sket{\psi_{\vb*{\eta}}},\hat{T}_s^\pidx{n}\sket{\psi_{\vb*{\eta}_n}}\Bigr).
\end{equation}

\subsection{Time-independent explicit case}

The equations above considerably simplify when considering time-independent Hamiltonians and an explicit integration method. Indeed, to any order $s$, we have 
\begin{equation}
	\hat{T}_s^\pidx{n} = \sum_{m=0}^{s-1}\lambda_m h_n^m\hat{\Pi}^{\pidx{n}m} = \sum_{m=0}^{s-1}\lambda_m (-i h_n \hat{H})^m \, ,
\end{equation}
with
\begin{equation}
	\lambda_m := \begin{cases}
		1, & m < 2,\\
		\vb*{b}^T\vb{a}^{m-2}\vb*{c}, & \text{else}.
	\end{cases}
\end{equation}
Two common Butcher tableaus corresponding to the fourth-order Runge-Kutta method are: 
\begin{equation}
	\renewcommand\arraystretch{1.2}
	\begin{array}
		{c|cccc}
		0 &  & & & \\
		1/2 & 1/2 & & & \\
		1/2 & 0 & 1/2 & & \\
		1 & 0 & 0 & 1 & \\
		\hline
		& 1/6 & 1/3 & 1/3 & 1/6 
	\end{array}
	\qquad\qquad
	\renewcommand\arraystretch{1.2}
	\begin{array}
		{c|cccc}
		0 &  & & & \\
		1/3 & 1/3 & & & \\
		2/3 & -1/3 & 1 & & \\
		1 & 1 & -1 & 1 & \\
		\hline
		& 1/8 & 3/8 & 3/8 & 1/8  \,
	\end{array}
\end{equation}
For these, we have:
\begin{table}[ht]
	\centering
	\label{tab:table1}
	\begin{tabular}{cccc}
		\toprule
		$\lambda_0$ & $\lambda_1$ & $\lambda_2$ & $\lambda_3$\\\midrule
		$1$ & $1$ & $1/2$ & $1/6$ \\  \bottomrule \, 
	\end{tabular}
\end{table}\\
This corresponds to the usual factor $\lambda_m = 1/m!$ of the truncated Taylor expansion of the propagator. Note, however, that this is generally no longer the case in a time-dependent scenario, as will appear below.

\subsection{Time-dependent implicit midpoint method}

The implicit midpoint method is characterized by the tableau
\begin{equation}
	\renewcommand\arraystretch{1.2}
	\begin{array}
		{c|c}
		1/2 & 1/2 \\
		\hline
		& 1
	\end{array}
\end{equation}
and yields
\begin{equation}
	\hat{T}_1^\pidx{n} = \hat{\mathds{1}} + h_n b_1\hat{\Phi}_{t_{n,1}} + h_n^2 a_{11}^2\hat{\Phi}_{t_{n,1}}^2 = \hat{\mathds{1}} - i h_n \hat{H}(t_n + h_n/2) - \frac{h_n^2}{4}\hat{H}^2(t_n + h_n/2).
\end{equation}

\subsection{Time-dependent Heun method}

The Butcher tableau of this explicit second-order Runge-Kutta method reads
\begin{equation}
	\renewcommand\arraystretch{1.2}
	\begin{array}
		{c|cc}
		0 &  & \\
		1 & 1 & \\
		\hline
		& 1/2 & 1/2 
	\end{array}
\end{equation}
and yields the following propagator:
\begin{align}
	\hat{T}_2^\pidx{n} &= \hat{\mathds{1}} + h_n b_1\hat{\Phi}_{t_{n,1}} + h_n b_2\hat{\Phi}_{t_{n,2}} + b_2 a_{21} h_n^2 \hat{\Phi}_{t_{n,2}}\hat{\Phi}_{t_{n,1}}\nonumber\\
	&= \hat{\mathds{1}} - i h_n \frac{\hat{H}(t_n) + \hat{H}(t_n + h_n)}{2} - \frac{h_n^2}{2}\hat{H}(t_n + h_n)\hat{H}(t_n).
\end{align}

\subsection{Time-dependent Ralston method}

The Butcher tableau of this explicit second-order Runge-Kutta method reads
\begin{equation}
	\renewcommand\arraystretch{1.2}
	\begin{array}
		{c|cc}
		0 &  & \\
		2/3 & 2/3 & \\
		\hline
		& 1/4 & 3/4 
	\end{array}
\end{equation}
and yields the following propagator:
\begin{align}
	\hat{T}_2^\pidx{n} &= \hat{\mathds{1}} + h_n b_1\hat{\Phi}_{t_{n,1}} + h_n b_2\hat{\Phi}_{t_{n,2}} + b_2 a_{21} h_n^2 \hat{\Phi}_{t_{n,2}}\hat{\Phi}_{t_{n,1}}\nonumber\\
	&= \hat{\mathds{1}} - i h_n \frac{\hat{H}(t_n) + 3\hat{H}(t_n + h_n)}{4} - \frac{h_n^2}{2}\hat{H}(t_n + h_n)\hat{H}(t_n).
\end{align}

\section{Scaling of the error with $\delta t$\label{app:e}}

The presented implicit midpoint method yields an update of the form
\begin{equation}
    \ket{\psi(t+\delta t)} = \hat{T}_\mathrm{mid}(t) \ket{\psi(t)} + O(\delta t^{3}),
\end{equation}
and as such may in principle be regarded as being of second order, provided one is able to approximate $\ket{\psi(t+\delta t)}$ with an NQS up to an error of order $O(\delta t^3)$. However, this condition cannot be satisfied with a normalized ansatz such as those used in this work. Indeed, the norm of the ideal updated state to be matched is given by:
\begin{equation}
    \ev{\hat{T}_{\mathrm{mid}}^\dagger\hat{T}_{\mathrm{mid}}^{\mathstrut}}{\psi(t)} = \ev{\Bigl(\hat{\mathds{1}} + \frac{3\delta t^2}{2}\hat{H}^2\Bigr)}{\psi(t)} + O(\delta t^3) = 1 + O(\delta t^2)\, .
\end{equation}
This must depart from $1$ to second order in $\delta t$, implying that the implicit midpoint method effectively yields a first-order update when using ansätze normalized by design.

In contrast to the midpoint method, the Heun method does not suffer from this issue, indeed:
\begin{align}\label{eq:Heun_tdagt}
    \hat{T}_{\mathrm{Heun}}^\dagger\hat{T}_{\mathrm{Heun}^{\mathstrut}} &= \Bigl( \hat{\mathds{1}} + i\delta t \hat{H} -\frac{\delta t^2}{2} \hat{H}^2\Bigr)\Bigl( \hat{\mathds{1}} - i\delta t \hat{H} -\frac{\delta t^2}{2} \hat{H}^2\Bigr) = \hat{\mathds{1}} + O(\delta t^4)\, .
\end{align}
This property allowed us to simplify the expression of the distance in Eq.~\eqref{eq:dist} as Eq.~\eqref{eq:simp_dist}, considerably reducing the complexity of evaluating the gradients of the loss function.

These considerations generalize to the time-dependent scenario, where for the Heun method we have
\begin{align}\label{eq:time_dep_norm}
    \hat{T}_{\mathrm{Heun}}^\dagger(t)\hat{T}_\mathrm{Heun}^{\mathstrut}(t) &= \Bigl(\hat{\mathds{1}} + i\delta t \hat{H}(t) - \frac{\delta t^2}{2} \hat{H}(t)\hat{H}(t+\delta t)\Bigr)\Bigl( \hat{\mathds{1}} - i\delta t \hat{H}(t) - \frac{\delta t^2}{2} \hat{H}(t+\delta t)\hat{H}(t)\Bigr)\nonumber \\
    &= \hat{\mathds{1}} + \delta t^2\hat{H}(t)^2 -\frac{\delta t^2}{2}\{\hat{H}(t), \hat{H}(t+\delta t) \} + O(\delta t^3).
\end{align}
Provided one can expand $\hat{H}(t+\delta t) = \hat{H}(t) + \delta t \partial_t \hat{H}(t) + O(\delta t^2)$, as is the case for any analytic quench, the second-order terms in Eq.~\eqref{eq:time_dep_norm} cancel out exactly, implying that the Heun method remains of second order when using normalized ansätze.

\end{document}